\documentclass[12pt]{article}
\usepackage{epsfig}
\usepackage{graphics}
\usepackage{dcolumn}
\usepackage{amsmath}
\usepackage{cite}
\usepackage{changebar}
\usepackage[english]{babel}      
\usepackage{color}
\definecolor{green}{rgb}{0.00,1.00,0.00}
\definecolor{blue}{rgb}{0.00,0.00,1.00}
\definecolor{gray}{rgb}{0.75,0.75,0.75}
\definecolor{red}{rgb}{1.00,0.00,0.00}
\definecolor{darkgreen}{rgb}{0.00,0.39,0.00}
\definecolor{darkgrey}{rgb}{0.66,0.66,0.66}
\definecolor{darkblue}{rgb}{0.00,0.00,0.55}
\definecolor{darkred}{rgb}{0.55,0.00,0.00}

\newcommand{\be}{\begin{equation}}
\newcommand{\ee}{\end{equation}}
\newcommand{\ba}{\begin{eqnarray}}
\newcommand{\ea}{\end{eqnarray}}

\begin{document}
\renewcommand{\figurename}{{\bf Fig.}}
\renewcommand{\tablename}{{\bf Tab.}}
\sloppy
\title{ 
  \vspace*{-2cm}{\normalsize\bf\hfill BI-TP 2004/12 \\ 
                                     \hfill IRB-TH-3/04} \\  
  \vspace*{1cm}
 Entropy for Color Superconductivity\\ 
 in Quark Matter
}

\author{David~E.~Miller$^{1,2,3}\,$\thanks{om0@psu.edu,
   \hspace*{2mm}  dmiller@physik.uni-bielefeld.de, \hspace*{2mm}
   dmiller@smith.irb.hr
}
 $\;$ and
 Abdel-Nasser~M.~Tawfik$^{1}\,$\thanks{tawfik@physik.uni-bielefeld.de}
 \\~\\ 
 {\small $^1$ Fakult\"at f\"ur Physik, Universit\"at Bielefeld, Postfach
 100131, D-33501
 Bielefeld, Germany} \\
 {\small $^2$ Theoretical Physics Division, Rudjer Boskovi\'c Institute,
   HR-10002 Zagreb, Croatia}\\   
 {\small $^3$  Department of Physics, Pennsylvania State University, 
         Hazleton, Pennsylvania 18201, USA} 
}

\date{}
\maketitle

\begin{abstract}
We study a model for color superconductivity with both three
colors and massless flavors including quark pairing. By using 
the Hamiltonian in the color-flavor basis we can calculate 
the quantum entropy. From this we are able to further investigate
the phases of the color superconductor, for which we find a rather
sharp transition to color superconductivity above a chemical potential 
around $290\;$MeV.
\end{abstract}

\section{\label{sec:1} Introduction}

At high quark chemical potentials and low temperatures 
the internal structure of hadronic matter has been conjectured 
to dissolve into a degenerate system of quarks. Such material
consisting of very cold dense quarks might exist in the interior of 
compact stellar objects. However, due to the difficulties of performing 
lattice simulations with high chemical potentials, it is still not 
possible to simulate the physics of these phases by using the 
usual lattice gauge computations. Nevertheless, 
a nonperturbative analysis at finite baryon density 
has been quite recently carried out on the lattice by using the 
\hbox{Nambu-Jona-Lasinio} model. The degenerate quarks 
near to the Fermi surfaces are generally expected to interact 
according to quantum chromodynamics (QCD) so that they can build up
Cooper pairs. This process may lead to superconducting quark
matter~\cite{Bar}.   

     Before we present the model for the quantum entropy $S$ in color
superconducting quark matter, one might well ask why $S$ as a
physical quantity is at all significant. 
There are a number of physical systems, which we have previously discussed 
in~\cite{Mill,MiTa1,MiTa2}, for which one has to  reconsider the meaning 
the third law of thermodynamics in its origin form. We have found that the
entropy remains finite even at absolute zero. However, if one were to take 
this fact into consideration, some aspects of the thermodynamical formulation 
would become more complicated. Our objective here is to interpret the reason
for the finiteness of this entropy at $T=0$ in the quark matter in relation 
to the quantum correlations in the ground state. How the presence of this 
finite quantum entropy term would affect the actual thermodynamics at
finite temperatures will be discussed elsewhere. The effects of 
including $S$ in the hadronic equation of state at low temperatures 
we have already studied in special cases~\cite{MiTa2}. 
To our knowledge, the works of Elze~\cite{Elze} have provided a start for 
addressing the question about the origin of the {\it entropy puzzle} in
the high-energy collisions. These works established a theoretical framework 
for discussing how two hadronic scattering initial states undergo hard 
collisions in quantum mechanically pure initial states. This situation 
can result in a high-multiplicity event corresponding to a highly impure 
thermal density matrix on the partonic level before hadronization. 
According to these works~\cite{Elze} the entropy is an unambiguous 
characteristic property of the quantum nature of the system. The
entropy production is clearly due to the {\it environmently induced 
quantum decoherence} in the observable subsystem. Therefore, we consider 
that there is no obvious theoretical reason to consider finite entropy 
for the interpretation of the particle multiplicity, while then explicitly 
setting its value to zero in other comparable cases. 
We have also used the prescription of von~Neumann for the entropy, 
which make use of the eigenvalues of the reduced density matrices. 
Thus we are able to give a first quantitative evaluation for the 
quarks' entropy inside the hadrons~\cite{Mill,MiTa1,MiTa2} 
under the condition that in the singlet state the quark groundstates 
are maximally mixed. In a previous work~\cite{MiTa3} we have given a
general evaluation of the entropy for the condensate in quark color
superconductivity using a nonrelativistic model based on the BCS theory . 

     In this work we start with an ultrarelativistic model Hamiltonian
for quarks with three colors and flavors which was proposed a few 
years ago~\cite{AlRaWi,Raj00}. In this framework we shall apply 
our previous calculations for the quantum ground state 
entropy~\cite{Mill,MiTa1,MiTa2} on these quarks under at large values
of the quark chemical potential. As mentioned above, the quantum entropy
means that entropy which  arises from quantum correlations from the quantum
fluctuations. These entities differ from the usual thermal fluctuations of
a statistical 
system in that they can also exists at zero temperature. In the above
mentioned model with three colors and flavors a total of nine mixed quark 
states are present. For such  mixed states in the pairs of two colored quarks
the quantum entropy can be expected to be temperature independent and 
equal to just $9\ln4$ in maximally mixed states~\cite{MiTa3}.

\section{\label{sec:2}Color Superconductor Model}

We now look more explicitly at the effects on the ground state in our model
with three colors and three flavors for the structure of the superconductivity.
In this part we shall calculate the quantum entropy from the gap equations
for the color superconducting state. For massless quarks the form for the 
Hamiltonian already has been written down for the three color-flavor
superconductivity~\cite{AlRaWi} arising from  quark-pairs at low temperatures
and high quark chemical potentials. It takes the following form:
\ba
H &=& \sum_{\rho,k>\mu}(k-\mu) a_{\rho}^{\dagger}({\bf k})
      a_{\rho}({\bf k}) +
      \sum_{\rho,k<\mu}(\mu-k) a_{\rho}^{\dagger}({\bf k}) 
      a_{\rho}({\bf k}) + 
      \sum_{\rho,{\bf k}} (k+\mu)b^{\dagger}_{\rho}({\bf k})b_{\rho}({\bf
        k}) \nonumber \\
  & & + \frac{1}{2} \sum_{\rho,{\bf p}} F(p)^2 Q_{\rho} e^{-i\phi({\bf p})}
  \left(a_{\rho}({\bf p}) a_{\rho}(-{\bf p}) + 
      b^{\dagger}_{\rho}({\bf p})b^{\dagger}_{\rho}(-\bf p) \right) \nonumber \\
  & & + \frac{1}{2} \sum_{\rho,{\bf p}} F(p)^2 Q_{\rho} e^{i\phi({\bf p})}
  \left(a^{\dagger}_{\rho}({\bf p}) a^{\dagger} _{\rho}(-{\bf p}) + 
      b_{\rho}({\bf p})b_{\rho}(-\bf p) \right). 
\ea
$F(p)^2$ is the form factor containing the cut-off $\Lambda$. 
$Q_{\rho}$ stands for the diagonalized form for the gap parameters, 
for  which $\rho = 1$ yields the color-flavor singlet gap-parameter
$\Delta_1$ and $\rho =2,\cdots,9$ result in the color-flavor gap
$\pm{\Delta_8}$.  
The first line of the Hamiltonian represents only the non-interacting parts, 
while the second and third lines thereof are the complex conjugate terms
of the interactions with opposite momenta. $a^{\dagger}_{\rho}$ and $a_{\rho}$ 
together with $b^{\dagger}_{\rho}$ and $b_{\rho}$ are the creation and
annihilation operators of the particle and antiparticle states, respectively. 
The index $\rho$, as given above, stands for both the color and flavor
degrees of freedom.
  
     For our present purpose we can treat $\rho$ in the same way as we had 
previously taken only the color degrees of freedom since the flavors 
now provide an exact symmetry in the limit of massless quarks. We take
$\mu$ as the quark chemical potential, for which all the momenta up to
$\mu=p_F$  have all the particle and antiparticle states completely 
occupied in the groundstate.

A proper parameterization for the annihilation and creation operators, 
respectively, had been already suggested~\cite{AlRaWi} as follows:
\ba
y_{\rho}({\bf k}) &=& \cos[\theta_{\rho}^y({\bf k})]a_{\rho}({\bf k}) +
  \sin[\theta_{\rho}^y({\bf k})] e^{i\xi_{\rho}^y({\bf k})}
  a_{\rho}^{\dagger}(-{\bf k})  \\ 
z_{\rho}({\bf k}) &=& \cos[\theta_{\rho}^z({\bf k})]b_{\rho}({\bf k}) +
  \sin[\theta_{\rho}^z({\bf k})] e^{i\xi_{\rho}^z({\bf k})}
  b_{\rho}^{\dagger}(-{\bf k}) \\  
y^{\dagger}_{\rho}({\bf k}) &=& \cos[\theta_{\rho}^y({\bf
  k})]a^{\dagger}_{\rho} 
({\bf k}) + \sin[\theta_{\rho}^y({\bf k})] e^{-i\xi_{\rho}^y({\bf k})}
  a_{\rho}(-{\bf k}) \\ 
z^{\dagger}_{\rho}({\bf k}) &=& \cos[\theta_{\rho}^z({\bf
  k})]b^{\dagger}_{\rho} 
({\bf k}) + \sin[\theta_{\rho}^z({\bf k})] e^{-i\xi_{\rho}^z({\bf k})}
  b_{\rho}(-{\bf k})   
\ea
Therefrom the following definitions are given:
\ba
\theta_{\rho}^y({\bf k}) &=& \frac{1}{2} \arccos
           \left(\frac{|k-\mu|}{\sqrt{(k-\mu)^2+F(k)^4Q_{\rho}^2}}\right)
            \\
\xi_{\rho}^y({\bf k}) &=& \phi({\bf k}) + \pi   \\
\theta_{\rho}^z({\bf k}) &=& \frac{1}{2} \arccos
           \left(\frac{|k+\mu|}{\sqrt{(k+\mu)^2+F(k)^4Q_{\rho}^2}}\right)
            \\
\xi_{\rho}^z({\bf k}) &=& -\phi({\bf k}) 
\ea
We may compare these complex expressions with the usual forms for the
Bogoliubov  transformations, from which we can write
\ba
u_{\rho}({\bf k})   &\equiv& \cos[\theta_{\rho}({\bf k})], \\
v_{\rho}({\bf k})   &\equiv& \sin[\theta_{\rho}({\bf k})] 
                             e^{i\xi_{\rho}({\bf k})}, 
\ea
Obviously, we get 
\ba
u^*_{\rho}({\bf k})u_{\rho}({\bf k}) + v^*_{\rho}({\bf k})v_{\rho}({\bf
  k})&=&1, \label{eq:cond}
\ea
which shows the canonical nature of these transformations.

\begin{figure}[thb]
\centerline{\includegraphics[width=12.cm]{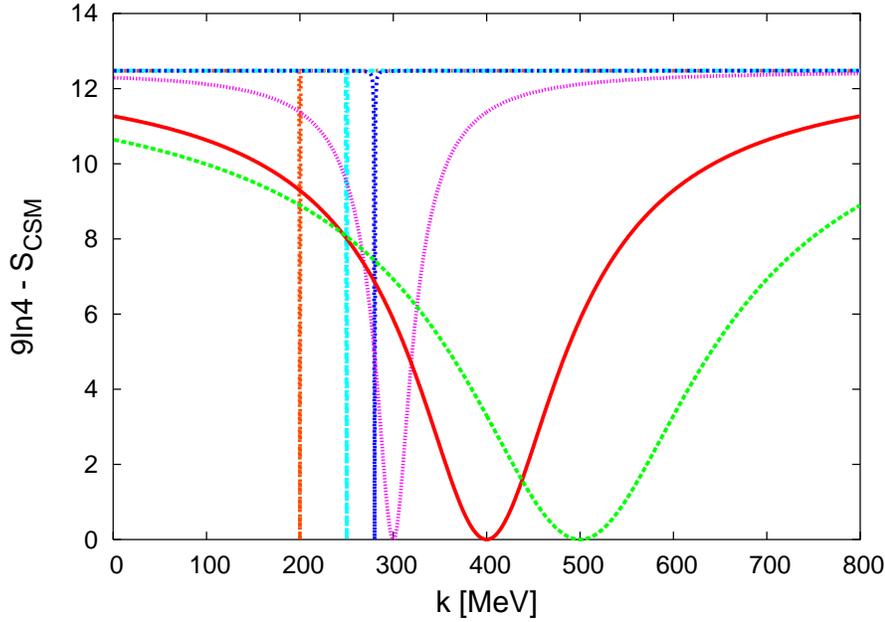}}
\caption{\it The momentum dependence of the difference
  between the maximum entropy in the ground state $9\ln4$  and 
  the entropy from the excitation of color superconductors is shown for
  different values of the chemical potential $\mu$ indicated by the
  zero-point on the graph.} 
  \label{fig:1}  
\end{figure}

\noindent
After we have carried out these canonical transformations, the 
form of the Hamiltonian for noninteracting quasiquarks takes on the quadratic 
canonical structure:
\ba
H &=& \sum_{{\bf k},\rho}
      \left[\left((k-\mu)^2+F(k)^4Q_{\rho}^2\right)^{1/2}
        y_{\rho}^{\dagger}({\bf k})y_{\rho}({\bf k})\right. \nonumber \\
  & & \left. \;\;\;\, + 
      \left((k+\mu)^2+F(k)^4Q_{\rho}^2\right)^{1/2} z_{\rho}^{\dagger}({\bf
        k})z_{\rho}({\bf k}) 
      \right]  
\ea
We now write out the following definitions:
\ba
\Upsilon^y_{\rho}({\bf k}) &\equiv& \frac{u^{{y}{*}}_{\rho}({\bf
    k})u^y_{\rho}({\bf k})} {v^{{y}{*}}_{\rho}({\bf k})v^y_{\rho}({\bf k})}
    = 
        \frac{\sqrt{(k-\mu)^2+
          F(k)^4Q_{\rho}^2} + |k-\mu|} {{\sqrt{(k-\mu)^2+
          F(k)^4Q_{\rho}^2}} - |k-\mu|} 
     \label{upsly1} \\
\Upsilon^z_{\rho}({\bf k}) &\equiv& \frac{u^{{z}{*}}_{\rho}({\bf
    k})u^z_{\rho}({\bf k})} {v^{{z}{*}}_{\rho}({\bf k})v^z_{\rho}({\bf k})}
    = 
       \frac{\sqrt{(k+\mu)^2+
          F(k)^4Q_{\rho}^2} + |k+\mu|} {\sqrt{(k+\mu)^2+
          F(k)^4Q_{\rho}^2} - |k+\mu|} 
\label{upslz1} 
\ea
\noindent          
For $k\neq 0$ we write the quantum entropy of entanglement for this color superconducting model in the following form:
\ba
S_{CSM} &=& g \;\sum_{\rho=1}^9
                 \left\{ \frac{\ln{\Upsilon^y_{\rho}({\bf
                        k})}}{\Upsilon^y_{\rho}({\bf k}) +1} 
                 + \ln\left(1+\frac{1}{\Upsilon^y_{\rho}({\bf k}}) \right)
                 + \right. \nonumber \\ 
        & & \hspace*{1.3cm}\left.\frac{\ln{\Upsilon^z_{\rho}({\bf
                k})}}{\Upsilon^z_{\rho}({\bf k})+1} 
                 + \ln\left(1+\frac{1}{\Upsilon^z_{\rho}({\bf k})}\right)
               \right\} \label{entopycolor1}
\ea
We first must set up the equations for the gaps $\Delta_1$ and $\Delta_8$,
which have been previously studied by Alford, Rajagopal and
Wilczek~\cite{AlRaWi}. 
Before we write down the complete gap equations, we briefly discuss the 
two color flavor superconducting model known as the 2SC Model. In structure
it is very similar to the simple nonrelativistic BCS model with the addition
of the antiparticle contribution. A not too great generalization of this type
of model leads to a two flavor and three color model~\cite{AlRaWi}, which
is the prior step to the above model. The 2SC phase is such that
the diquarks condense while the chiral symmetry is restored. It has a simple 
equation for the gap $\Delta$  similar to the above BCS Model. We write the
gap equation in the form~\cite{Raj00}
\ba
1 & = &\frac{2G}{V}\sum_p \left\{\frac{1}{\sqrt{(p-\mu)^2+\Delta^2}}  
       +\frac{1}{\sqrt{(p+\mu)^2+\Delta^2}}\right\}. \label{coup1}
\ea
We can convert the sum over all the momenta p into an integral 
${\mathcal{I}}[\Delta]$ so that
\ba
{\mathcal{I}}[\Delta]& = &\frac{G}{\pi^2}\int^{\Lambda}_0 k^2dk\left(\frac{1}
{\sqrt{(k-\mu)^2+\Delta^2}} +\frac{1}{\sqrt{(k+\mu)^2+\Delta^2}}\right)
\label{coup2} 
\ea
This form of the gap equation
\ba
1 & = &{\mathcal{I}}[\Delta] \label{gapeq1}
\ea
can be evaluated and substituted in the above
equation for the entropy. However, it does not properly reflect the fully
extended color-flavor symmetry of our above model Hamiltonian. Nevertheless,
we can use this simpler equation with the constituent quark mass $M$ replacing 
the gap parameter at vanishing chemical potential in order to determine the 
coupling $G$. The coupled integral equations for the 
gaps represent the singlet and octet decomposition of this extended
symmetry~\cite{AlRaWi}. In our evaluation of the equations for the gaps
$\Delta_1$ and  
$\Delta_8$ we use the step-function cut-off with $\Lambda=800\;$MeV. Thus we
write the two gap equations~\cite{Raj00} in the following form:
\ba 
 \Delta_1 &=& -2{\Delta_8}{\mathcal{I}}[\Delta_8]  \label{delt1} \\
 \Delta_8 &=& -{\Delta_1} \left(1 + {\mathcal{I}}[\Delta_1] \right)/4
\label{delt8} 
\ea
We use these gap equations for $\Delta_1$ and $\Delta_8$ in order to obtain
$\Upsilon^y_{\rho}({\bf k})$ and $\Upsilon^z_{\rho}({\bf k})$ with $F(k)=1$ for
$k < \Lambda$ and zero above. These quantities are substituted into the above
equation for the quantum entropy $S_{CSM}$. The results for $9\ln4 - S_{CSM}$ 
are shown in Fig.~\ref{fig:1} for different values of the quark chemical
potential $\mu$. We can see the effects of the gaps  $\Delta_1$ and
$\Delta_8$ on the quantum entropy. Below a critical value of the chemical
potential around $290\;$MeV the gaps vanish so that the spread of $S_{CSM}$
also vanishes. We see this in a very narrow line that extends downwards to
zero at values of $\mu$ under $290\;$MeV.\\

\begin{figure}[thb]
\centerline{\includegraphics[width=12.cm]{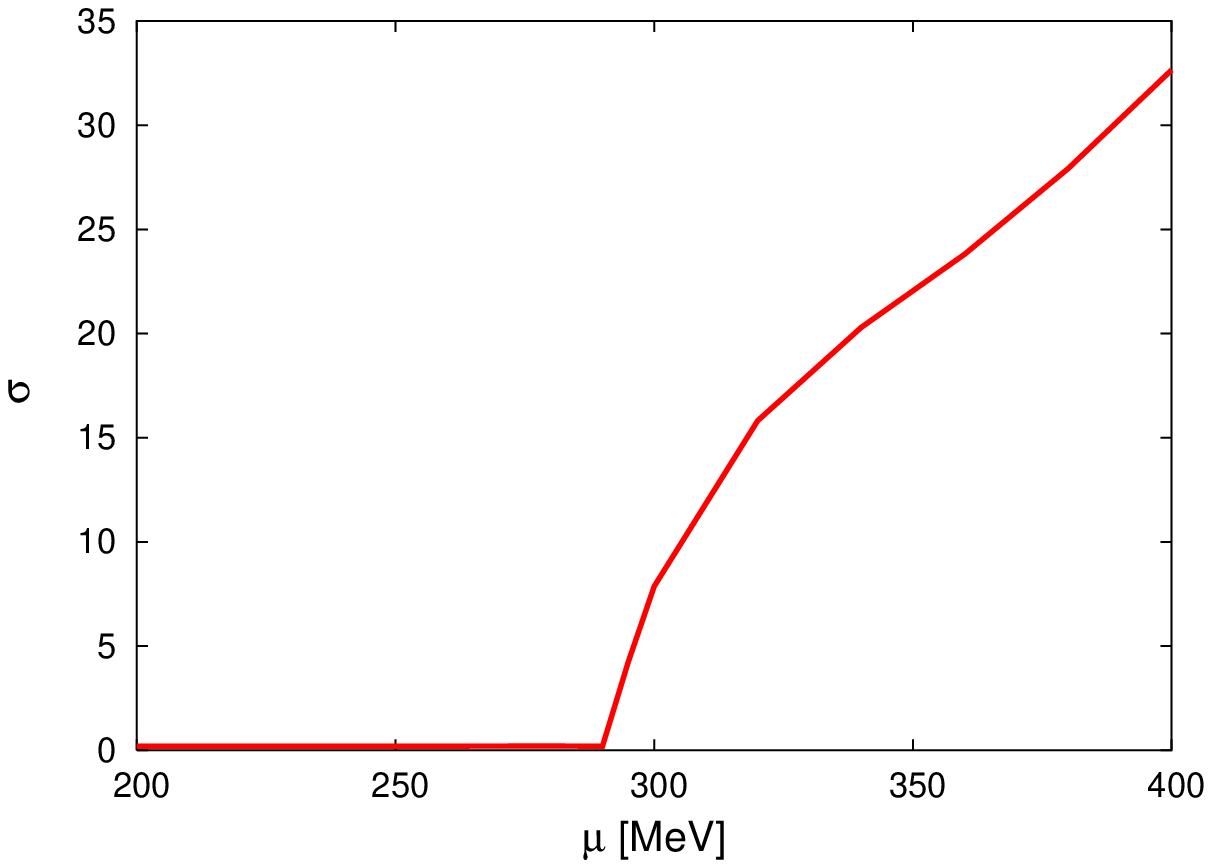}}
\caption{\it The dispersion $\sigma$ taken from the distributions given in  
  Fig.~\ref{fig:1} plotted against the corresponding $\mu$
  value. $\sigma\neq0$ for quark chemical potential \hbox{$\mu>290\;$MeV}.} 
  \label{fig:2}  
\end{figure}

\section{\label{sec:3}Results and discussion}

Now we discuss the results from the computation of the quantum entropy
$S_{CSM}$ derived in the equation~\ref{entopycolor1}.
Figure~\ref{fig:1} shows the difference between the entropy from the quark 
pairing in color superconductors Eq.~\ref{entopycolor1} and the maximum
value of the groundstate entropy for nine states given by $~9\ln4~$.  
The dependence of $~9\ln4-S_{CSM}~$ upon the momenta $k$ is plotted for 
different values of the chemical potential $\mu$. This difference has 
a zero-point when the value of the momentum $~k=\mu~$. In this figure 
we notice for the zero-point values above  $\mu=290$ that there is always 
a finite spread in the curve around the zero-point. From this fact 
we can extract at the halfheight value the fullwidth $\Gamma$. 
By means of a direct  comparison with the Gaussian distribution 
we take the dispersion to be the standard deviation $\sigma$. 
In this case we have simply the fullwidth $\Gamma=2\sqrt{2\ln2}\;\sigma$ 
at half maximum. Then we can compute the dispersion  
$\sigma$ from the distributions given in the next figure~\ref{fig:2} 
as a function of the corresponding $\mu$  value. In this model with
three massless flavors taken together with the three colors from the
usual $SU(3)_c$ we are able to analyze the dependence of the gaps $\Delta_1$  
and $\Delta_8$ on the quark chemical potential $\mu$ using the physical
quantity $S_{CSM}$.\\
  
\section{\label{sec:4}Conclusion}
Finally we can conclude that the quantum entropy is a good indicator of
the phase transition arising from the presence of the two gaps 
$\Delta_1$ and $\Delta_8$. As we have seen in the figure~\ref{fig:2},
the dispersion  $\sigma$ becomes finite at a chemical potential of about 
$290\;$MeV, which is very near the previously calculated value of around
$300\;$MeV for the gaps appearing in the CFL phase~\cite{AlRaWi,Raj00}.
Further discussion of these models and their properties relating to the
phenomena of color superconductivity has appeared quite 
recently~\cite{Bowers1}. In our present work we have shown that by
using the entropy $S_{CSM}$ we have an additional quantity which
can be computed to show the transition to color superconductivity.
Its deviation from the pairing value is due to the two gaps. Furthermore, 
the dispersion $\sigma$ relates directly  to the correlations,
which,in this case, these correlations are between the quarks 
within the pairs, which is similar to the diquark structure.  
These ideas could lead to further investigations relating to 
the expected correlations between the diquarks as well as, perhaps, 
find future applications in the theoretical studies of the recently 
experimentally discovered pentaquark states~\cite{Penta}. These states
can be interpreted~\cite{JaWi} as a bound state of four quarks and an
antiquark, which consist of two highly correlated quark pairs. \\
~\\
~\\
{\bf Acknowledgments}\\
In preparing this work we have benefited from many stimulating discussions with
Krzysztof~Redlich. We acknowledge the further helpful discussions with 
David~Blaschke. D.E.M. would like to thank Tome Anti\'ci\'c and 
Kresko Kadija for useful information on the NA49 experiment 
relating to the various pentaquark states. 
He is very grateful to the support from the Pennsylvania State 
University Hazleton for the sabbatical leave of absence and
to the Fakult\"at f\"ur Physik der Universit\"at Bielefeld, 
especially to Frithjof Karsch. He also would like to thank the
Fulbright Scholar Program  and the Ministry of Science, Education and Sport
of the Republic of Croatia for the support at the Ruder Boskovi\'c Institute. 
~\\

\end{document}